\documentstyle[11pt,newpasp,twoside,psfig]{article}
\begin{document}

\title{Global versus Nuclear Starbursts }
\author{Fran\c{c}oise Combes }
\affil{DEMIRM, Observatoire de Paris, 61 Av. de l'Observatoire,
F-75 014, Paris, France}

\begin{abstract}
The strongest starbursts are observed towards galaxy nuclei,
or circumnuclear regions. However in interacting galaxies,
star formation is also triggered in overlap regions far from
nuclei, in spiral arms and sometimes in tidal tails.
What is the relative importance of  these starbursts?
What kind of starformation is dominating, as a function of redshift?
These different starbursts occur in different dynamical
conditions (global and local): gravitational instabilities,
density waves, radial flows, shear, cloud collisions, density accumulations,
and they have been investigated with the help of numerical simulations.
Gravitational instabilities are necessary to initiate star formation, but they
are not sufficient; galactic disks are self-regulated through these instabilities
to have their Toomre Q parameter of the order of 1, and thus this criterium
is in practice unable to predict the onset of intense star formation.
Super star clusters are a characteristic SF mode in starbursts, and
might be due to the rapid formation of large gas complexes.
Star formation can propagate radially inwards, due to gravity torques 
and gas inflow, but also outwards, due to superwinds, and energy outflows:
both expanding or collapsing waves are observed in circumnuclear regions. 
Mergers are more efficient in forming stars at high redshift, because
of larger gas content, and shorter dynamical times.
The relation between nuclear starbursts and nuclear activity 
is based on the same fueling mechanisms, but also on
reciprocal triggering and regulations. 
\end{abstract}

\section{Observations : where are starbursts located ?}

It is a widely observed fact that starbursts are concentrated
in nuclei, and in particular the strongest ones (ULIRGs).
But there can be exceptions, such as:

\begin{itemize}
\item the Antennae, Arp 299, where star formation
is more intense in overlap regions between the two galaxies,
\item in bright spiral arms (like M51, etc..)
\item the Cartwheel and other collisional ring galaxies: the starburst occurs
in the ring, sometimes in the nucleus, or toward the second developing ring,
\item in nuclear resonant rings of barred galaxies; this ring might shrink with time
and the starburst drifts towards the center, as seems to be the case in 
M82 : a fossil region M82B NE (de Grijs et al 2001), has been 
studied 1kpc from the central nuclear starburst.
\end{itemize}

\begin{figure}[t]
\centerline{
\psfig{figure=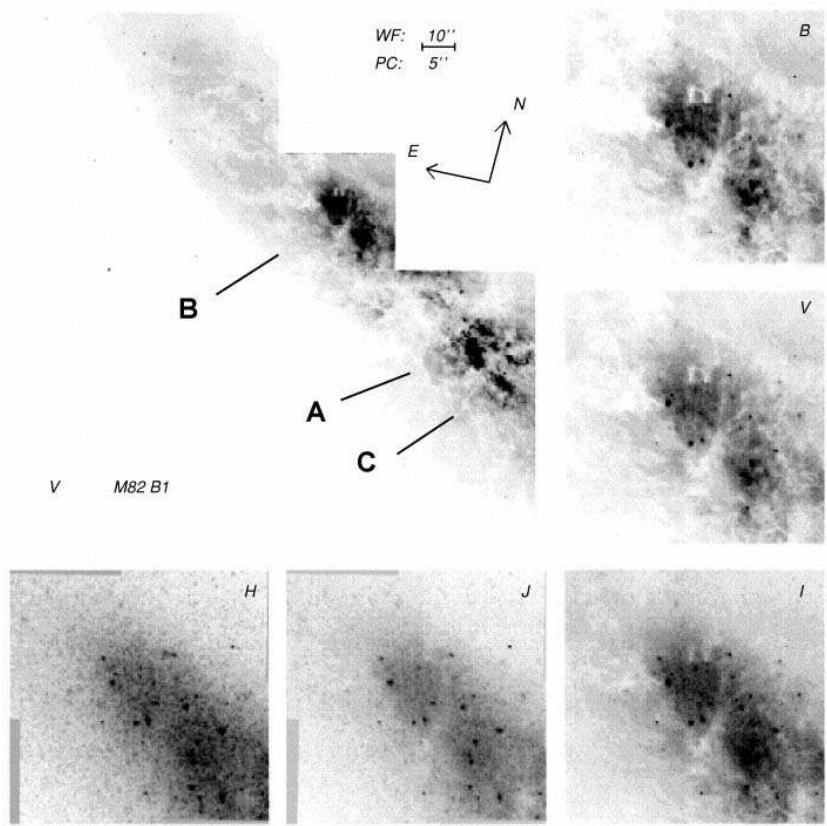,width=11cm,angle=0}
}
\caption{HST-WFPC2 image (V-band) of M82 (in the center), and
PC-field images in B, V, I and NICMOS in J \& H, from de Grijs et al. (2001).
The PC and  NICMOS images are centered on the fossil starburst
(region B), while regions A and C indicate the present on-going
nuclear starburst. In M82B, a large system of evolved super star clusters
has been found.}
\label{fig1}
\end{figure}

M82 is a good opportunity to study the evolution of starburst location:
in the M82B fossil region (see figure 1), stars formed 
100 Myr ago, with a comparable amplitude than the present starburst 
in the center. De Grijs et al. (2001) find there an important (113) number of 
evolved super star clusters (SSC). Their detailed age study conclude that the
starburst begun 2 Gyr ago, with a peak 600 Myr ago, and stopped about 
30Myr from now. This episode could coincide to a previous 
passage/interaction of the companion M81. The evolution of the SSCs
is compatible with them being progenitors of globular clusters.

This evolution of the starburst location could correspond
to ring concentration and ring evolution. Indeed, in
barred galaxies star formation is frequently in nuclear rings
(cf Buta et al 2000, NGC 1326; Maoz et al 2001, 
NGC 1512, NGC 5248), and in particular bright knots in 
the rings.

Sometimes star formation can occur even farther
from the center: in tidal dwarfs (e.g. Duc et al. 2000), shells, garlands,
large HII complexes in the outer regions, as in M101 or NGC 628 
(Leli\`evre \& Roy 2000).
Or the nucleus does not concentrate the star formation activity,
which is more randomly distributed, as in dwarf irregulars.
A recently studied example is NGC 4214 (Beck et al. 2000; MacKenty et al. 2000),
where interferometric CO observations (Walter et al 2001) reveal
that the star formation is not always coinciding with the 
gaseous concentrations. If one CO complex is indeed the site
of a starburst, a comparable one, at the same distance from the
center, is completely quiecent.

An interesting question is to estimate the relative importance
of starburst, and more quiescent or ``steady-state'' star formation
 in the global rate of star formation of the Universe.
If a  starburst is defined as having a rate larger than 50 M$_\odot$/yr,
an estimation from NICMOS images in the Hubble Deep Field
conclude that both processus appear similar in 
importance (Thompson  2000).

\section{Dynamical mechanisms}

 Since the fuel for star formation is the interstellar gas, it is
straightforward to assume that the star formation rate should
be proportional to some power of the volumic gas density in galaxies,
as done by Schmidt (1959). Following this assumption, Schmidt
derived that this power should be around 
n=2\footnote{Schmidt compared the scale-height of
the galactic plane in gas and young stars to derive this power, and
found a high value because the molecular hydrogen distribution
was not known at that time}
in the solar neighborhood.
However, this local hypothesis has revealed very difficult to confirm, although
there is of course some correlations between global gas density
and star formation rate in a Galaxy. The difficulty is certainly 
related to the time delays and time-scales for star formation processes
and subsequent feedback, and also to the fact that the gas can be
stabilised by dynamical mechanisms, instead of forming stars.

\subsection{Global statistical studies}

So far, only global quantities have been correlated 
with success, when the gas surface density is averaged
out over the whole galaxy, and the same for the star formation rate.
The star formation tracer can vary, from the H$\alpha$ flux
for normal galaxies, to the Far Infrared luminosity L(FIR) for
starbursts, which are highly obscured (Kennicutt 1998).

While the starbursts explore a wider range and dynamics
of parameters, the relation between 
the global gas surface density and star formation rate (SFR) 
is the same for extreme and normal galaxies: it is possible
to derive a ``global'' Schmidt law, with a power n=1.4 (Kennicutt 1998).
$$
\Sigma_{SFR} \propto  \Sigma_{gas}^{1.4}
$$
(cf figure 2).
Another formulation works as well
$$
\Sigma_{SFR} \propto  \Sigma_{gas} \Omega \propto \Sigma_{gas}/t_{dyn}
$$
where $\Omega$ is the angular frequency in the galaxy, which is 
inversely proportional to the dynamical time-scale $t_{dyn}$.

Several justifications can be found a posteriori: if the star formation
is locally due to the gravitational instability of the gas, this occurs on
a free-fall time-scale, and the star-formation rate is:
$$
d\rho_*/dt = \rho_{SFR} \propto \rho_{gas}/\tau_{ff} \propto  \rho_{gas}^{1.5}
$$
very close to the power n=1.4; but the correlation is not observed locally.
Globally, this applies also, if the star formation is due to the global
gravitational instability of the gas disk, that occurs in a dynamical time-scale:
$$
d\Sigma_*/dt = \Sigma_{SFR} \propto \Sigma_{gas}/t_{dyn} 
$$
which accounts for the second formulation. Alternatively, star formation
could be triggered in marginally stable clouds, by the crossing of
spiral arms, and the frequency of arm crossing is
proportional to  $\Omega - \Omega_p$  (Wyse \& Silk 1989),
or roughly to  $\Omega$ far from corotation (where the clouds
never cross the arms).

This second formulation might also explain the Tully Fisher relation
(Silk, 1997; Tan 2000), since if
$\Sigma_{SFR} \propto  \Sigma_{gas} \Omega$, then
$$
L_b  \propto \Sigma_{SFR} R^2 \propto \Sigma_{gas} v_{circ} R
$$
with $v_{circ}^2 \propto \Sigma R$ from the virial, and 
provided that  $\Sigma_{gas} \propto \Sigma_*$ is verified over the
main spiral classes (Roberts \& Haynes 1994),
it can be deduced that $L_b  \propto v_{circ}^3$.

The numerical values found for the global Schmidt law correspond
to an SFR of 10\% of gas per orbit transformed into stars,
at the outer edge typically for normal galaxies.
The much higher SFR in starbursts could be only a consequence of their much 
higher surface density: indeed $\Sigma_{gas}$ is observed
to be 100 to 10 000 higher, and the star formation efficiencies (SFE)
about  6-40 times higher.
This higher efficiency can also be attributed to
smaller dynamical time-scales, since starbursts usually happen in nuclear regions.

A starburst is obtained as soon as dynamical mechanisms have brought 
gas to the center; this can occur through  gravity torques on dynamical 
time-scales. The gas infall must be sufficiently rapid to overcome the
feedback processes, that will blow the gas out. These processes, such
as supernovae explosions and violent stellar winds, occur on time-scales
of 10$^7$ yr, the life-time of O-B stars. The latter is unchanged at any galactic
radius, being intrinsic to stellar physics. Only in nuclei dynamical torques can 
bring the gas faster than these feedback mechanisms.

\bigskip

The global statistical studies appear to be slightly different for extreme 
starbursts (Taniguchi \& Ohyama 1998).
The exponent of the global Schmidt law is more near n=1, and
$\Sigma_{SFR} \propto \Sigma_{gas}$
(cf also Young et al 1986). As for radial distribution, 
there is no correlation between $\Sigma_{FIR}$ and $\Sigma_{gas}$.
 It is the total gas amount of a galaxy that governs the infrared
luminosity L(FIR).

\begin{figure}[t]
\centerline{
\psfig{figure=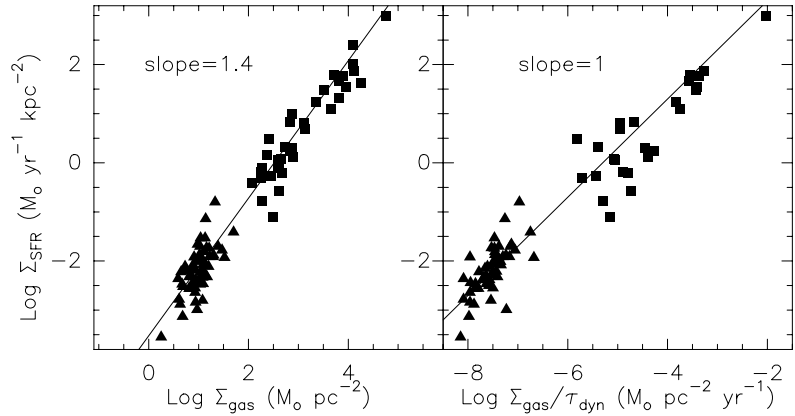,width=11cm,angle=-90}
}
\caption{ Relations between Star Formation Rate (SFR) and gas surface
densities, from Kennicutt (1998). {\it Left}: a global ``Schmidt'' law,
with a slope of n=1.4 as the best fit. {\it Right}: linear relation between
SFR and gas over dynamical time ($\tau_{dyn}$) ratio. Normal galaxies are the
triangles, while starburst galaxies are the squares. }
\label{fig2}
\end{figure}

\subsection{Parameters governing the SFR}

The difficulty is that there are many physical parameters determining 
the SFR and SFE in galaxy disks. Along the Hubble sequence, the star
formation rate increases towards late-type, wich could be due 
to dynamical instability increasing with decreasing bulge-to-disk ratio.
The SFE has been found to decrease with size (Young 1999). However,
this could be due to a metallicity effect, since SFE is computed from 
the infrared to H$_2$ ratio, SFE= L(FIR)/M(H$_2$), itself derived
from L(FIR)/L(CO), and L(CO) could lead to an underestimation
of H$_2$ in low-mass under-abundant galaxies.

The SFR also depends on environment, since galaxy interactions
are one of the most widely recognized trigger of starbursts.
Gravity torques are also essential for radial gas flows, and
thus the bar phase or chronology might play a role, as well as the gas
content.

The most essential physical parameters are :
\begin{itemize}
\item  Gravitational instability, the main trigger of star formation;
this might explain the existence of a threshold of gas density for star formation,
the critical surface density $\Sigma_c$ (Quirk 1972, Kennicutt 1989);

\item  Cloud-cloud collisions, a process also 
proportional to a power of local volumic density,  $\propto \rho^2$,
that could imply  a local Schmidt law. It is possible to account for observations of 
SFR and SFE, by considering only collisions (Scoville 2000, Tan 2000);

\item  Tidal forces;
interaction and mergers are the main trigger of starbursts
(e.g. Kennicutt et al. 1987, Sanders \& Mirabel 1996). 
Also, after the interaction, the binary black hole thus formed
can trigger a starburst by its dynamical perturbations (Taniguchi \& Wada 1996);

\item  Gas density (Schmidt 1959, Kennicutt 1998), and radial gas flows 
due to gravity torques (bars);

\item  Supernovae/winds can also drive star formation
(Wada \& Norman, 1999, 2001); clouds marginally stable could be
driven into gravitational instability by an
excess of pressure, a blast wave (Koo \& McKee 92, Heckman et al 96, 
Taniguchi et al 98). Star formation can be contagious, since
it propagates local instabilities.
\end{itemize}

Since all these phenomena play a role in the star formation, a
global Schmidt law, averaged over the whole galaxy, is not sufficient to
disentangle the relative importance of each process. 
In particular, local studies reveal that the
 gas density alone is not a sufficient parameter to predict SFR and SFE
(cf gas concentrations without starbursts, Jogee \& Kenney 2000).

It is tempting to test the stability of gaseous disks, with the Toomre criterium
Q (and its equivalent formulation as a critical gas density $\Sigma_c$),
in order to explain the occurence of star formation in special regions
or  galaxies. However in a dynamical time, gravitational instabilities
are able to heat a disk until the stability criterium is verified in 
almost all disks, and external parameters are not included in the
criterium.

\subsection{Why are Q and $\Sigma_c$ actually not very useful 
to predict star formation trigger and starburst activity?}

The main problem is that the criterium for gravitational instabilities
is often undissociated from the criterium of star formation.
But in reality, if gravitational instabilities are necessary for
star formation, they are not sufficient. There are still some
other parameters that are essential, controlling the onset of
star formation in a gas medium that has formed self-gravitating
structures, and those parameters are still to be sorted out and
quantified to build a criterium for star formation:

\smallskip
{\it 1) Self-regulation }

Gravitational instabilities are so important that disks are
self-regulated to have the Toomre Q parameter of the order of 1.
Indeed, as soon as Q falls below 1 because of gas dissipation,
the disk becomes gravitationnally unstable: these instabilities
have for immediate effect to increase the velocity dispersion, 
and heat the disk so that Q $\sim$ O(1) again (e.g. Lin \& Pringle
1987).
But this self-gravitating process occurs even in the absence of
star formation, so that Q $\sim$ O(1) in any disk and cannot help
to predict star formation.

For instance, in the outer parts of spiral disks, where it is obvious
that there is no star formation at all, the HI gas is observed to be
gravitationally unstable and form structures at all scales: there are
spiral arms, giant clumps, and a mass spectrum of clouds
(structures down to the smallest structure possible to see with the present
21cm beams). It is therefore likely that Q is also there of the order
of 1, and the disk self-regulated.
This occurs also inside some irregular galaxies, possessing a lot of gas,
without star forming activity like N2915 (Bureau et al 1999). The gas
has developped gravitational instabilities, spiral arms, etc..

\bigskip
{\it 2) Multi-phase gas and multi-components stability}

Let us emphasize that Q and $\Sigma_c$ characterizing
the stability of disks, should be computed taking into account
all components, gas and stars, and in case of several gas components,
the total multi-phase medium. This is not possible analytically, since
the different components have not the same velocity dispersion, but an 
empirical criterium has been proposed as:
$$
\frac{1}{Q} = \frac{1}{Q_{gas}} + \frac{1}{Q_{star}}
$$
(Jog 1992).
Therefore, each component has a weight $\propto \Sigma/\sigma$ in
the stability. 
When there is a large surface density of gas, the Q$_{gas}$ term 
dominates, and it is justified to compute Q and $\Sigma_c$ taking
gas only into consideration. But as soon as the gas surface density
is depleted for some reason (for instance inside circum-nuclear rings),
or the gas is heated ($\sigma_{gas}$ increases), then the Q$_{star}$ 
term has to be taken into account, ensuring that the total Q is always
of the order of 1, over the whole disk (e.g. Bottema 1993). 
As for the gas velocity dispersion, when there exist
non-axisymmetric features, like spiral arms, bars, etc.. the corresponding
streaming motions have to be included in $\sigma_{gas}$ (which does
not reduce to the local sound speed velocity of the order of 10km/s),
since it is precisely these streaming motions that are the consequences
of disk heating by spiral waves and gravitational instabilities.

\bigskip
{\it 3) Spatial averaging scale}

A problem in estimating Q and $\Sigma_c$ is also the scale at which they
are averaged, and the results can change completely according to the 
spatial resolution of the observations.
We know that the interstellar medium is fractal and possess structures at all scales,
from 100 pc to $\sim$ 10 AU. The gas surface density increases towards 
small scales, by about 1-2 orders of magnitude; the critical surface density
might be reached or not, according to the spatial scale of averaging
(Klessen 1997; Wada \& Norman 1999, 2001; 
Semelin \& Combes 2000; Huber \& Pfenniger 2001).

\bigskip
{\it 4) Uncertainty on the H$_2$ gas density}

The biggest uncertainty in computing the gas surface density is
the CO to N(H$_2$) conversion ratio. This ratio can vary within a factor
2 or 10, according to metallicity, CO excitation, temperature, density, etc..
(Rubio et al 1993, Taylor et al 1998, Combes 2000), 
and since Q $\sim$ O(1) in galaxy disks
anyway, it is quite impossible to ascertain that Q is larger or smaller than 1
if such systematic uncertainties are
attributed to the gas density. Due to the latter, 
it is likely that systematics will find $\Sigma_{gas} < \Sigma_c$ 
for non-star forming regions, where the CO is not excited (or the metallicity
not enough), and $\Sigma_{gas} > \Sigma_c$ for starbursts ($^{12}$C is a
primary element, and the abundance of CO is enhanced in starbursts).

\bigskip
{\it 5) Intermittency}

Star formation can be inhibited or triggered by other phenomena, such as
supernovae, stellar winds, external or internal wave triggers and this does
not enter the Q and $\Sigma_c$ estimations. In a nuclear disk, simulations
by Wada \& Norman (2001), the density undergoes phases of episodic 
and recurrent star formation (of the order of $\sim$ 10 Myr periodicity),
and the estimation of $\Sigma_c$ are the same for periods with and
without star formation. Here is introduced a hidden parameter, which 
is the past history of star formation. A galactic disk region might be
quiescent, only in between two star formation phases for instance.

\bigskip

In conclusion, gravitational instabilities ensure that all spiral disks have
Q $\sim$ O(1) at all radii: the gas component is structured in clouds that
are marginally stable. Only transiently the disk can be brought out of 
equilibrium. Only a sudden trigger is necessary to start a
starburst, and these are difficult to recognize. This could be a sudden
radial gas flow due to a bar, or the tidal action of a companion, that
strengthens or creates a bar, that will bring gas to the nucleus,
when the dynamical time-scale is short.

To have a starburst, gas must be gathered in a very short time-scale, smaller
than $\sim$ 10$^7$ yr, shorter than the onset of feedback from the first
OB stars formed, through supernovae explosions and stellar winds, before
the starburst can blow the gas out.
In nuclei, the dynamical time-scale is shorter, while the feedback time-scale
is constant all over the disk (being intrinsic to the life-time of OB stars).
That might explain why starbursts are always more conspicuous in nuclei.

The original Schmidt law is a local one, and involves the volumic density
$\rho$. At this stage, one should consider that the surface density in inner and outer
parts of the galaxies have not the same weight for gravitational 
instabilities, because of the flaring of gas and star densities towards
the outer parts.

\subsection{Influence of bars}

It was found, with IRAS fluxes as a tracer of star formation, that barred galaxies
were more frequently starbursting (Hawarden et al. 1986), and had also
more radio-continuum central emission, attributed to star formation 
(Puxley et al. 1988). From a statistical sample of more than 200 starbursts and normal 
galaxies, Arsenault (1989) found a much larger frequency of barred and
ringed types among the starbursts, suggesting 
that active formation of stars in the nuclei of spirals is
linked to the perturbation of bars and gravity torques.

But such a correlation is not without any controversy: Pompea
\& Rieke (1990) do not find that strong bars appear
an absolute requirement for high infrared luminosity in isolated galaxies.
Markarian starbursting galaxies are less barred than a sample of normal
galaxies (Coziol et al. 2000).

At least the molecular gas appears much more concentrated in barred galaxies
(Sakamoto et al 1999), which is expected form gravity torques.
This gas concentration should trigger nuclear starbursts, according
to the Schmidt law.

As for nuclear activity itself, the correlation between the presence 
of bars and AGN activity is presently unclear, as described already in this 
conference.
Peletier et al (1999) and Knapen et al (2000) 
have shown that Seyferts have more bars than normal galaxies 
(results at 2.5$\sigma$).
Seyferts have curiously a lower fraction of strong bars 
(Shlosman et al 2000), perhaps pointing toward 
the destruction of bars by massive black holes.
Besides, Seyferts have more outer rings, by a factor 3 or 4 
(Hunt \& Malkan 1999). Since the outer rings are the vestiges of 
the action of bars, this supports the scenario of bar destruction by
central gas accretion and massive black holes.

An interesting feature recently discovered in stellar kinematics
of star-forming galaxies with an active nucleus, is the drop in velocity dispersion
in the central kpc. This was found thanks to ISAAC on the VLT (Emsellem et al 2000,
and this conference). This drop is unexpected, especially since the
dispersion should increase towards the massive black hole. But the phenomenon
can be transient, and due to kinematically cold stars just formed from the gas
in the nuclear disk fueled by the bar torques.

As for numerical simulations, starbursts are easily reproduced,
in particular triggered by galaxy interactions and mergers. The star-formation
is due to radial gas flows, driven by the bars formed in the interaction
(e.g. Mihos \& Hernquist 1994, 96; Bekki 1999). The bar is thus central to
the starburst. The presence of a bulge, which has a
stabilising influence on disks against bar formation, is determinant
in the occurence of the starburst. In galaxies with a large bulge-to-disk 
ratio, the intense starburst has to wait the merging, and the final 
gas infall, while galaxies without large bulges undergo 
repetitive starbursts.

Other dynamical perturbations, like lopsidedness and $m=1$ waves
are also triggering starbursts: in this case, star formation is mainly in the disk,
and not boosted in the nucleus (Rudnick et al 2000).

\section{Large Gas Complexes and Stellar Clusters}

Due to the large gas surface density in nuclear
starbursts, the critical length for self-gravity in the disk center
(the scale with the largest growth rate) is also very large:
$$
\lambda_{crit} = 4 \pi^2 G \Sigma / \kappa^2
$$
where $\Sigma \sim \Sigma_{gas}$, since the gas
is dominating there. The corresponding self-gravitating 
mass is $\lambda^2 \Sigma$, or $\propto \Sigma^3$.
 Figure 3 gives orders of magnitude for these values,
typical sizes and masses 200 pc, 10$^9$ M$_\odot$.

\begin{figure}[t]
\centerline{
\psfig{figure=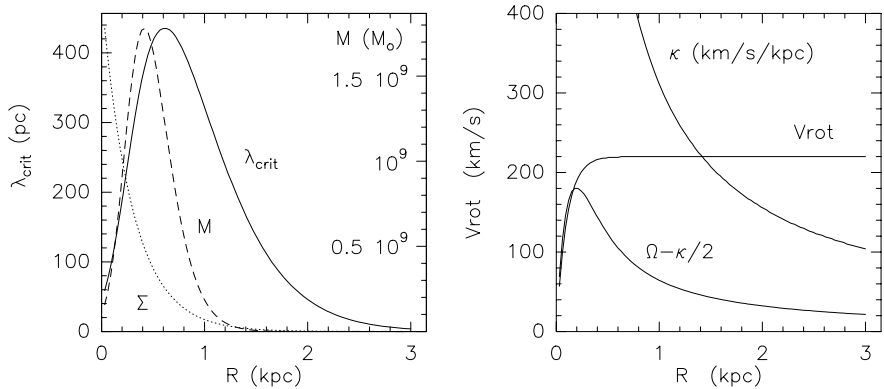,width=12cm,angle=-90}
}
\caption{{\bf Left} Critical length scale for self-gravitation $\lambda_{crit}$
(solid line) as a function of radius, together with the corresponding mass of structures
(dashed, right scale), for an exponential gas surface density, as represented
in dotted line (maximum $\Sigma_0$ = 5 10$^3$ M$_\odot$/pc$^2$ or 
3 10$^{23}$ H$_2$ cm$^{-2}$). {\bf Right} The critical length scale has been computed 
for this rotation curve, and corresponding frequencies $\kappa$, and
$\Omega-\kappa/2$.
}
\label{fig3}
\end{figure}

These super complexes will collapse, and may form
super star clusters, if another factor is tuned,
the time-scale before feedback effects
come into play, and regulate the star formation.
The collapse of gas must be sudden enough
(in $<$10 Myr), so that OB stars and SN cannot limit
the process. This means that the free-fall time is short
enough, and therefore that the volumic density is larger
than 2 M$_\odot$/pc$^3$. This is indeed verified for the
typical masses and sizes determined above, but not for usual giant
molecular clouds.

\bigskip
 
Another point of view to see the
formation of these large complexes, is to introduce the velocity dispersion
(Elmegreen et al. 1993). In interacting and merging galaxies, one
characteristic is that the tidal perturbations have increased
velocity dispersion above that of a quiescent disk, and 
the corresponding pressure stabilises locally the gas up to a larger
Jeans length. The complexes that form are then bigger.

The largest growth rate for instabilities in the disk occurs at 
the scale $\lambda_{crit}$ considered above, which is
also equal to the Jeans length:
$$
\lambda = \sigma^2/G \Sigma
$$
since the Toomre parameter Q $\sim \frac{\sigma \kappa}{\pi G \Sigma}  \sim$ 1.
In fact, the kinetic pressure stabilises all scales below Jeans length,
and the galactic rotation stabilises all scales above $\lambda_{crit}$,
the equality between the two ensuring the disk stability. If the disk
is slightly out of equilibrium, it is those common scales that are
unstable more quickly.

With this second formulation, the mass of the complexes are
proportional to $\sigma^4/\Sigma$, and grow at a rate $\tau_{ff} = \sigma/\Sigma$
showing the large importance of velocity dispersion.

\bigskip

Super Star Clusters (SSC) are
young star clusters of extraordinary luminosity and compactness.
They are one of the dominant modes of star formation in starbursts, and they
are thought to be a formation mechanism for globular clusters.
A major breakthrough from HST has been to show that
globular clusters form still at the present time, through starbursts
(e.g. Schweizer 2001). 
The question has been raised of the SSC contribution to the total luminosity:
it appears only moderate in ULIRGs (Surace et al 1998).
In Arp 220 for example (Shioya et al 2001), there are
three conspicuous nuclear SSC (galactic radius $<$0.5kpc),
which correspond to about 0.2 L$_{tot}$ (they are 
heavily obscured $>$ 10 mag). The 
disk SSC (0.5 $<$ radius $<$ 2.5 kpc), of lower luminosity, 
represent a negligible contribution.
SSC also form in starbursting dwarfs, with properties
quite similar to larger interacting/merging galaxies (e.g Telles 2001).
In these systems, they could represent a significant part of the 
luminosity. Their formation is thought to be triggered by the high pressure
experienced by the gas complexes in a starburst environment.

\section{Feedback, regulation, propagation}

The study of stellar populations, through multiband photometry
and spectroscopy, together with HII regions
and molecular gas distribution, and assisted by starburst
evolutionary models, leads to the determination of the age
and history of the star formation in a galaxy disk. 
It is possible to constrain the IMF, often found to be
biased towards high-masses in starbursts, and to follow the 
propagation of the starburst radially.

In some cases, the star formation propagates 
inside out, a good example being the ring of NGC 1614
(Alonso-Herrero et al 2001): here a nuclear starburst is
identified within 45pc, surrounded by a ring of HII regions
of 600 pc in diameter, tracing a younger burst.
These HII regions, about 10 times the intensity of 30 Doradus,
lie inside a ring of molecular gas, as if the star formation
wave was propagating radially outwards.
 
In the LINER galaxy NGC 5005, Sakamoto et al (2000) 
identify a stream of molecular gas, linking the inner ring
of the bar to the nuclear disk, likely to correspond to the ILR. 
This stream represents a high rate of bar-driven inflow and 
they suggest that a major fueling event is in progress in 
this galaxy. The gas flow could then be episodic rather than continuous.
Recurrent starbursts are then expected.

In other cases, the star formation appears to propagate
outside in: older star formation in a disk/ring of 200pc
in diameter surrounds a younger nuclear starburst in NGC 6764
(Schinnerer et al 2000):
two starbursts with decay times of 3 Myr occurred 3-5 Myr and 15 to 50 Myr ago.
However, a constant star formation scenario over 1 Gyr 
(at a rate of 0.3 M$_\odot$/yr) could also explain the data.

The ringed barred galaxy NGC 4314 also supports the outside in scenario:
a ring of dense molecular gas is observed inside the radio-continuum ring  
(Combes et al 1992; Benedict et al 1996; Kenney et al 1998).
The gas ring, inside the nuclear hot spots, evolves slowly, reducing 
its radius due to friction exerted by the background stars on the 
giant molecular clouds.  

This shrinking ring of star formation
is expected from the dynamical evolution of the gaseous nuclear ring.
Alternatively, feedback processes from violent star formation,
such as supernovae, bipolar gas outflows, etc... are expected to compress
the surrounding gas outwards, and to trigger star formation inside out.

\section{Starbursts as a function of redshift}

\subsection{More efficient star formation at high z}

It is now widely recognized that starbursts were
more frequent in the past, and galaxy imaging  at
high redshift with the Hubble Space Telescope has revealed
considerable evolution. Although there are still many
systematic biases in high-z studies, it appears that
galaxies were more numerous, and in particular
more perturbed and irregular. The Hubble classification is
difficult to pursue at high z. Galaxies are knotty, have  
less organised structures, and much less bars (van den Bergh et al 2001).
Their irregular appearance can be attributed to interactions, since
there are more pairs and more mergers  at high redshift
(Lefevre et al 2000).

The higher star formation rate at high z is easy to explain :
\begin{itemize}
\item More gas at high  redshift
\item Higher interaction and merging rates
\item Dynamical time shorter (to accrete gas)
\end{itemize}

In the frame of the hierarchical scenario, where large galaxies
today have been formed by succesive mergers of smaller entities,
the first haloes to form at high redshift have very small masses.
 But they are also denser, because they virialise from a much denser
universe, due to expansion. The volumic density is going as 
$(1 + z)^3$, and the dynamical time-scale inside these haloes
is going as  $\tau_{dyn} \sim (1 + z)^{-3/2}$.  Therefore,
in addition to the larger fraction of mergers at $z=2$,
the efficiency for a given merger to form stars is even higher.
The feedback mechanism, related to the life-time of OB stars,
has no reason to vary with redshift, and the time-scale to accrete
gas is shorter at high $z$. 

Also, it is easy to predict, since galaxies accumulate mass in
their bulge through secular evolution and galaxy interaction/merger,
that galaxies in the past were more unstable, having a smaller
bulge-to-disk ratio. Bar instability is then more violent,
with more gas accretion, and bars are destroyed also more
quickly. The fact that bars are transient might explain the 
observed lower bar frequency, although the present observations
are still preliminary.

\subsection{Relation between starburst and AGN }

Starbursts and AGN compete for gas fuel. They relie on the 
same dynamical mechanisms to be feeded and active. 
The main consequence of radial gas flow due to bars and
gravity torques is not only a nuclear starburst and an AGN, but
also the bulge growth, and a massive black hole growth.
However, the amount of gas required to grow the BH over Gyrs is small,
ensuring that both can occur simultaneously, which is reflected in the
observed correlation between the final masses: 
M$_{BH}$ = 0.1-0.2\% M$_{bulge}$
(Magorrian et al 1998, Ferrarese \& Merritt 2000). 
The relation between starbursts and AGN is not only circumstancial,
but there are effective regulation from one to the other and 
reciprocally. For instance, the central BH mass can modify
the  central dynamics, so as to favor gas accretion, or instead
to destroy a bar, and stop accretion and star formation.
Nuclear starbursts produce outflows (such as M82, N253) 
that regulate the BH grow, while the compact stellar clusters formed can provide
fuel to the BH through stellar mass loss (e.g. Norman \& Scoville 1988).

Although there is a massive black hole in almost every galaxy today,
most of them are quiescent. According to quasars counts and
luminosity as a function of redshift (e.g. Boyle et al 1991),
QSOs were more numerous and more powerful in the past.
This means that those black holes that were active were more massive,
while at low redshift, only more modest black holes are entering 
their activity cycle (Haehnelt \& Rees 1993).

We can deduce that the AGN-starburst connection at high 
redshift was a little different than today: composite objects
were more dominated by their AGN, due to their greater
black hole mass. 

Another point comes from their lower bulge mass: the inner
Lindblad resonance was less frequent, and in this case
the gas can be accreted all the way down to the nucleus, 
since it is not stalled at ILR. Of course, the time-scale
of gas accretion is longer when there is no resonance, but this
might be compensated by the shorter dynamical time-scale.
It is then likely that a black hole was easier to feed at high $z$.
Besides, the accretion being easier, the regulating mechanism
was operating faster, then destroying the bar after a shorter
time-scale.  All these phenomena have to be tackled
in details, to determine their actual effect on evolution.

\section{Conclusions} 

The detailed processes leading to star formation at large scales
in a galaxy disk or in the nucleus are still not well known. 
Many physical mechanisms can explain observations:
gravitational instabilities, cloud-cloud collisions, 
density-wave and radial flows, propagating
star-formation, galaxy interactions...

Empirical laws like the "global" Schmidt law do not help
in disentangling the role of all these physical phenomena.
Moreover, a "local" Schmidt law is still an unconfirmed
paradigm, since there is
no tight correlation between local gas density and SFR
density.

The main factor towards giant starbursts is the quick flow
of gas in a concentrated region in a short enough time-scale ($<$10 Myr),
to beat the stellar feedback processes.
This can only be provided by gravity torques in gaseous disks
(due for instance to galaxy interactions, that trigger bars, etc...)

This mechanism might be preponderant only at late Hubble times, 
when galaxies are massive, with stabilising bulges.
At earlier times (z$>$ 1), galaxies are less evolved and less concentrated;
 they are not stabilised against gravitational instabilities;
those can be violent, triggering spontaneous bursts, with a chaotic appearance,
accounting for the irregular and knotty images observed at high redshift.

Starbursts and AGN are often observed in symbiosis in galaxies,
they are not only fed by the same mechanisms, but sometimes
regulate each other. 
The observations at high redshift help to get insight in the
time evolution of both, leading to parallel growth of bulges
and supermassive black holes. 
Dark haloes forming earlier are denser,
explaining why supermassive black holes
forming earlier are more massive (Haehnelt \& Kauffmann 2000).

Evolutionary cosmological models (N-body simulations + semi-analytical 
experiments) succeed to some extent to reproduce observations: 
they use a local Schmidt law for star formation 
$$
d\rho_*/dt = c_* \rho_{gas}/max (t_{cool}, t_{dyn})
$$
and introduce schematically the stellar feedback, by yielding 
energy at each star formation to increase the bulk motion of the gas. 
Simulations retrieve rather well the
slope of the Tully-Fischer relation, which appears to be
not very sensitive to SF prescriptions
 (e.g. Steinmetz \& Navarro 2000).
But there is a big problem to retrieve the zero point:
at a given rotational velocity, model galaxies are 2 magnitude dimmer
than observed galaxies. The problem is now well
identified, the dark matter is too much concentrated in the models, 
and there is not enough baryons in the central regions of a galaxy disk.
 This has also been remarked in fitting rotation
curves and in particular 
of dwarf irregulars, that are dominated by dark matter.
This is independent of cosmological parameters (CDM, or $\Lambda$CDM),
although the efficiency to transform baryons into stars is much higher
in CDM ($\sim$ 100\%) than in $\Lambda$CDM ($\sim$ 40\%) 
(Navarro \& Steinmetz 2000). 

\acknowledgements
 I am very grateful to Johan Knappen and collaborators for 
the organisation of such a pleasant and
fruitful conference, to their sponsors and the CNRS for their financial help.


\begin{references}
\reference Alonso-Herrero A., Engelbracht C.W., Rieke M.J., Rieke G.H., Quillen A.C.:
    2001, ApJ 546, 952  
\reference Arsenault, R.: 1989, A\&A 217, 66
\reference Beck S., Turner J., Kovo O.: 2000, AJ 120, 244  
\reference Bekki K.: 1999, ApJ  510, L15
\reference Benedict, F. G., Smith, B. J., Kenney, J. D. P.: 1996, AJ 111, 1861
\reference Bottema, R.: 1993, A\&A 275, 16
\reference Boyle B.J., Jones L.R., Shanks T. et al.: 1991, in ``The Space Density of
  Quasars'', ASP Vol 21, p. 191
\reference Bureau, M., Freeman, K. C.,
    Pfitzner, D. W., Meurer, G. R.: 1999, AJ 118, 2158
\reference Buta R., Treuthardt P., Byrd G., Crocker D.: 2000, AJ 120, 1289 
\reference Combes, F., Gerin, M., Nakai, N., Kawabe, R., Shaw, M. A.: 1992, A\&A 259, L27
\reference Combes F.: 2000, in ``Molecular Hydrogen in Space'',
   eds. F. Combes \& G. Pineau des For\^ets, Cambridge Univ. Press, p. 275
\reference Coziol, R., Considère, S., Davoust, E., Contini, T.: 2000, A\&A 356, 102
\reference de Grijs R., O'Connell R., Gallagher J.: 2001, AJ 121, 768  
\reference Duc, P.-A., Brinks, E., Springel, V., Pichardo, B.,
    Weilbacher, P., Mirabel, I. F.: 2000, AJ 120, 1238
\reference  Elmegreen B., Kaufman M., Thomasson M.: 1993, ApJ  412, 90
\reference  Emsellem E., Greusard, D., Combes F., Friedli D., Leon S., P\'econtal E.,
     Wozniak H.: 2001, A\&A 368, 52
\reference Ferrarese L., Merritt D.: 2000, ApJ 539, L9
\reference  Haehnelt M.G., Kauffmann G: 2000, MNRAS 318, L35
\reference  Haehnelt M.G., Rees M.J.: 1993, MNRAS 263, 168
\reference Hawarden, T. G., Mountain, C. M., Leggett S.K., Puxley, P. J.: 1986, MNRAS 221, P41
\reference Heckman, T. M., Dahlem, M., Eales, S. A., Fabbiano, G., Weaver, K.: 1996,
   ApJ 457, 616   
\reference Huber D., Pfenniger D.: 2001, A\&A preprint (astro-ph/0105501)
\reference  Hunt L.K., Malkan M.A.: 1999 ApJ  516, 660
\reference Jog, C.: 1992, ApJ 390, 378
\reference Jogee S. \& Kenney J.D.P.: 2000, in ``Dynamics of Galaxies: from the Early
  Universe to the Present'', eds F. Combes, G. Mamon, V. Charmandaris,
  ASP Conf, Vol. 197, p. 193
\reference Kenney, J. D. P., Benedict, G. F., Friedli, D.: 1998, BAAS, 193, 53.03
\reference Kennicutt, R. C.,  Roettiger, K. A., Keel, W. C., van der Hulst, J. M.,
  Hummel, E.: 1987, AJ 93, 1011
\reference Kennicutt R.C.: 1989, ApJ 344, 685
\reference Kennicutt R.C.: 1998, ApJ 498, 541
\reference Klessen R.: 1997, MNRAS 292, 11
\reference Knapen, J. H., Shlosman, I., Peletier R.F.: 2000, ApJ 529, 93
\reference Koo B-C, McKee C.F.: 1992, ApJ 388, 93 \& 103
\reference Lefevre O.,  Abraham, R., Lilly, S. J., et al.: 2000, MNRAS 311, 565
\reference Leli\`evre M., Roy J-R.: 2000, AJ 120, 1306
\reference Lin D.N.C., Pringle J.E.: 1987, MNRAS 225, 607
\reference MacKenty J.W., Maiz-Apellaniz J., Pickens C., Norman C., Walborn N.: 2000
  AJ 120, 3007   
\reference Magorrian, J., Tremaine, S., Richstone, D., et al.: 1998, AJ 115, 2285
\reference Maoz D., Barth A., Ho L.C., Sternberg A., Filipenko A.: 2000, BAAS, 197, 78.02
\reference Mihos J. C., Hernquist L.: 1994, ApJ 437, 611
\reference Mihos, J.C., Hernquist, L.:1996, ApJ, 464,  641
\reference Navarro J., Steinmetz M.: 2000, ApJ 538, 477
\reference Norman C., Scoville N.Z.: 1988, ApJ 332, 124
\reference Peletier, R. F., Knapen, J. H., Shlosman, I., et al.: 1999, ApJS 125, 363
\reference Pompea S., Rieke G.: 1990, ApJ 356, 416
\reference Puxley, P. J., Hawarden, T. G., Mountain, C. M.:1988, MNRAS 231, 465
\reference Quirk W.J.: 1972, ApJ 176, L9
\reference Roberts M., Haynes M.: 1994, ARAA 32, 115
\reference Rubio M., Lequeux J., Boulanger F.: 1993, A\&A 271, 9
\reference Rudnick G., Rix H-W., Kennicutt R.: 2000, ApJ 538, 569  
\reference Sakamoto, K., Okumura, S. K., Ishizuki, S., Scoville, N. Z.: 1999, ApJ
   525, 691
\reference Sakamoto, K., Baker, A.J., Scoville, N. Z.: 2000, ApJ 533, 149
\reference Sanders D., Mirabel F.: 1996, ARAA 34, 749
\reference Schinnerer E., Eckart A., Boller T.: 2000, ApJ 545, 205 
\reference Schmidt, M.: 1959, ApJ 129, 243
\reference Schweizer F.: 2001, in ``Extragalactic Star Clusters'', eds. E. Grebel,
   D. Geisler, D. Minniti, IAU 207 (astro-ph/0106345)
\reference Scoville N.Z.: 2000, in ``Dynamics of Galaxies: from the Early
  Universe to the Present'', eds F. Combes, G. Mamon, V. Charmandaris,
  ASP Conf, Vol. 197, p. 301 
\reference Semelin B., Combes F.: 2000, A\&A 360, 1096
\reference Shioya Y., Taniguchi Y., Trentham N.: 2001, MNRAS 321, 11  
\reference Shlosman, I., Peletier, R. F., Knapen, J. H. : 2000, ApJ 535, L83
\reference Silk J.: 1997, ApJ 481, 703 
\reference Steinmetz M. \& Navarro J.: 2000, in ``Dynamics of Galaxies: from the Early
  Universe to the Present'', eds F. Combes, G. Mamon, V. Charmandaris,
  ASP Conf, Vol. 197, p. 165 
\reference Surace, J. A., Sanders, D. B., Vacca, W, D., Veilleux, S., Mazzarella, J. M.: 1998,
   ApJ 492, 116
\reference Tan J.: 2000, ApJ 536, 173
\reference Taniguchi Y., Wada K.: 1996, ApJ 469, 581  
\reference Taniguchi Y., Ohyama Y.: 1998, ApJ 509, L89  
\reference Taniguchi Y., Trentham N., Shioya Y.: 1998, ApJ 504, L79
\reference Taylor C.L., Kobulnicky H.A., Skillman E.D.: 1998, AJ 116, 2746
\reference Telles E.: 2001, in ``Extragalactic Star Clusters'', eds. E. Grebel,
   D. Geisler, D. Minniti, IAU 207 (astro-ph/0106341)
\reference Thompson R.: 2000, BAAS, 197, 117.03
\reference van den Bergh S., Cohen J., Crabbe C.: 2001, ApJ, preprint (astro-ph/0104458)
\reference Wada K., Norman C.: 1999, ApJ 516, L13
\reference Wada K., Norman C.: 2001, ApJ 547, 172
\reference Walter F., Taylor C., H\"uttemeister S., Scoville N., 
       McIntyre V.: 2001, AJ 121, 727 
\reference Wyse R., Silk J.: 1989, ApJ 339, 700
\reference Young J.S., Kenney, J. D., Tacconi, L., Claussen, M. J.,
       Huang, Y.-L., Tacconi-Garman, L., Xie, S., Schloerb, F. P.: 1986, ApJ 311, L17
\reference Young J.S.: 1999, ApJ 514, L87
\end{references}
\end{document}